\documentstyle[12pt]{article}

\newcommand{\eq}{\begin{equation}}
\newcommand{\en}{\end{equation}}

\newcommand{\eqa}{\begin{eqnarray}}
\newcommand{\ena}{\end{eqnarray}}
\newcommand{\eqan}{\begin{eqnarray*}}
\newcommand{\enan}{\end{eqnarray*}}

\newcommand{\lbl}{\label}

\newcommand{\sect}[1]{\setcounter{equation}{0}\section{#1}}

\newcommand{\CMP}[1]{Comm. Math. Phys.\ {\bf #1}\ }

\newcommand{\NP}[1]{Nucl. Phys.\ {\bf #1}\ }

\newcommand{\PR}[1]{Phys. Rev\ {\bf #1}\ }

\newcommand{\RMP}[1]{Rev. Mod. Phys.\ {\bf #1}\ }

\def\sqr#1#2{{\vcenter{\hrule height.#2pt
     \hbox{\vrule width.#2pt height#1pt \kern#1pt
        \vrule width.#2pt}
     \hrule height.#2pt}}}

\def\thinspace{\kern .16667em}

\def\Dir{\nabla\kern-2ex\Big{/}}

\def\reali{{\hbox{\s@ l\kern-.5ex R}}}
\def\naturali{{\hbox{\s@ l\kern-.5ex N}}}
\def\interi{{\mathchoice
 {\hbox{Z\kern-1.5mm Z}}
 {\hbox{Z\kern-1.5mm Z}}
 {\hbox{{Z\kern-1.2mm Z}}}
 {\hbox{{Z\kern-1.2mm Z}}}  }}

\def\unity{{\hbox{\s@ 1\kern-.8mm l}}}
\def\uno{{\hbox{ 1\kern-.8mm l}}}
\def\part{\partial}

\def\aa{\alpha}
\def\bb{\beta}
\def\cc{\chi}

\def\dd{\delta}

\def\ee{\epsilon}

\def\vf{\varphi}
\def\gg{\gamma}
\def\GG{\Gamma}

\def\ll{\lambda}

\def\pp{\psi}
\def\PP{\Psi}
\def\pb{\bar\psi}
\def\PB{\bar\Psi}
\def\rr{\rho}

\def\ss{\sigma}

\def\tt{\theta}

\begin{document}
\begin{titlepage}
\begin{flushright}
NORDITA-93-41\\
June 1993\\
hep-th/9306091
\end{flushright}
\vspace*{0.5cm}
\begin{center}
{\bf
\begin{Large}
{\bf
THE MASTER FIELD OF QCD$_2$ AND THE 'T HOOFT EQUATION
\\}
\end{Large}
}
\vspace*{1.5cm}
         {{\large Marco Cavicchi}\footnote{Also:
          Istituto di Fisica ``Augusto Righi'',
          Via Irnerio 46, 40100 Bologna,
          Italy. E-mail: cavicchi@bo.infn.it ~/ ~ cavicchi@nbivax.nbi.dk },
          {\large Paolo Di Vecchia}}
         \\[.3cm]
          NORDITA\\
          Blegdamsvej 17, DK-2100 Copenhagen \O \\
          Denmark\\
\vspace*{.5cm}
         {\large Igor Pesando}\footnote{E-mail PESANDO@NBIVAX.NBI.DK,
22105::PESANDO, 31890::I\_PESANDO}
         \\[.3cm]
          The Niels Bohr Institute\\
          University of Copenhagen\\
          Blegdamsvej 17, DK-2100 Copenhagen \O \\
          Denmark
\end{center}
\vspace*{0.7cm}
\begin{abstract}
{
We rewrite the action for $QCD_2$ in the light cone gauge only in terms
of a bilocal mesonic field. In this formalism the $1/N$
expansion can be done in a straightforward way by  a saddle point
technique that determines the master field to be identified with the
vacuum expectation value of the bilocal field. Finally we show that the
equation of motion for the fluctuations around the master field is
identical with the 't Hooft meson equation.
}
\end{abstract}
\vfill
\end{titlepage}

\setcounter{footnote}{0}
\def\ut{{\tilde u}}
\def\zt{{\tilde z}}
\def\dz{{\sqrt{2}z}}

\def\uij{U_{ij}}
\def\ucij{U^\dagger_{ij}}
\def\uji{U_{ji}}
\def\ucji{U^\dagger_{ji}}
\def\dag{\dagger}
\def\V{{\cal V}}
\def\S{{\cal S}}
\def\zpm{z_{\pm}}
\def\zp{z_+}
\def\zm{z_-}
\def\ddt{{\dd T}}
\def\mucr{\mu_{cr}}

\newcommand{\mat}[4]{\left(
                     \begin{array}{cc}
                     {#1} & {#2} \\
                     {#3} & {#4}
                     \end{array}
                     \right)
                    }
\newcommand{\vett}[2]{\left(
                      \begin{array}{c}
                     {#1} \\
                     {#2}
                     \end{array}
                     \right)
                    }
\newcommand{\vet}[2]{\left(
                     \begin{array}{cc}
                     {#1} &  {#2}
                     \end{array}
                     \right)
                    }
\newcommand{\ft}[3]{\int {d^{#1}{#2}\over (2\pi)^{#1}} ~ e^{i {#2}\cdot{#3}} }

\def\rhm{\rho_-}
\def\rhp{\rho_+}
\def\sgm{\sigma_-}
\def\sgp{\sigma_+}

\def\rd{\sqrt{2}}
\def\usrd{{1\over\sqrt{2}}}
\def\dxy{\delta^2(x-y)}
\def\dij{\delta^{ij}}
\def\dsi{\partial_{x^-}}
\def\dta{\partial_{x^+}}

\newcommand\modu[1]{|{#1}|}

\newcommand\psiind[3]{\psi^{#1 #2}_{#3}}
\newcommand\psibind[3]{{\bar\psi}^{#1 #2}_{#3}}
\newcommand\psiindd[2]{\psi^{#1 #2}}
\newcommand\psibindd[2]{{\bar\psi}^{#1 #2}}

\def\pbai{\psibindd{A}{i}}
\def\pbaj{\psibindd{A}{j}}
\def\pbbi{\psibindd{B}{i}}
\def\pbbj{\psibindd{B}{j}}
\def\pai{\psiindd{A}{i}}
\def\paj{\psiindd{A}{j}}
\def\pbi{\psiindd{B}{i}}
\def\pbj{\psiindd{B}{j}}

\def\paip{\psiind{A}{i}{+}}
\def\pajp{\psiind{A}{j}{+}}
\def\pbip{\psiind{B}{i}{+}}
\def\pbjp{\psiind{B}{j}{+}}
\def\paim{\psiind{A}{i}{-}}
\def\pajm{\psiind{A}{j}{-}}
\def\pbim{\psiind{B}{i}{-}}
\def\pbjm{\psiind{B}{j}{-}}

\def\pbaip{\psibind{A}{i}{+}}
\def\pbajp{\psibind{A}{j}{+}}
\def\pbbip{\psibind{B}{i}{+}}
\def\pbbjp{\psibind{B}{j}{+}}
\def\pbaim{\psibind{A}{i}{-}}
\def\pbajm{\psibind{A}{j}{-}}
\def\pbbim{\psibind{B}{i}{-}}
\def\pbbjm{\psibind{B}{j}{-}}

\newcommand\fai[4]{{#1}^{{#2}}_{{#3}} ({#4}) }

\def\rjimyx{\fai{\rho}{j i}{-}{y,x}}
\def\rilmxz{\fai{\rho}{i l}{-}{x,z}}
\def\rljmxy{\fai{\rho}{l j}{-}{x,y}}
\def\rljmzy{\fai{\rho}{l j}{-}{z,y}}
\def\rijpxy{\fai{\rho}{i j}{+}{x,y}}
\def\rljpxy{\fai{\rho}{l j}{+}{x,y}}
\def\rijmxy{\fai{\rho}{i j}{-}{x,y}}
\def\rljmxy{\fai{\rho}{l j}{-}{x,y}}
\def\rllmuu{\fai{\rho}{l l}{-}{u,u}}

\def\sijmxy{\fai{\ss}{i j}{L}{x,y}}
\def\sljmxy{\fai{\ss}{l j}{L}{x,y}}
\def\sljmzy{\fai{\ss}{l j}{L}{z,y}}

\def\sijpxy{\fai{\ss}{i j}{R}{x,y}}
\def\sljpxy{\fai{\ss}{l j}{R}{x,y}}
\def\sljpzy{\fai{\ss}{l j}{R}{z,y}}

\def\aijp{{\alpha^{i j}_+}}
\def\aijm{{\alpha^{i j}_-}}
\def\bijp{{\beta ^{i j}_L}}
\def\bijm{{\beta ^{i j}_R}}

\def\aijmxy{\fai{\aa}{i j}{-}{x,y}}
\def\aijpxy{\fai{\aa}{i j}{+}{x,y}}
\def\ajimyx{\fai{\aa}{j i}{-}{y,x}}
\def\ajipyx{\fai{\aa}{j i}{+}{y,x}}

\def\bijpxy{\fai{\bb}{i j}{L}{x,y}}
\def\bijmxy{\fai{\bb}{i j}{R}{x,y}}
\def\bjipyx{\fai{\bb}{j i}{L}{y,x}}
\def\bjimyx{\fai{\bb}{j i}{R}{y,x}}
\def \bigd{{\cal D}}

\sect{Introduction}
The large $N$ colour expansion~\cite{HOOFT1} is probably the most promising
technique for
arriving at an analytical understanding of long distance properties of non
abelian gauge theories. The major obstacle to the realization of this
program is the fact that it has been impossible up to now to compute the
sum over planar diagrams in a matrix model in a closed form unless one
works in zero or one dimension~\cite{PARISI}.

Recently Kazakov and Migdal~\cite{KM} have introduced a scalar matrix model
coupled to a gauge field without its kinetic term that on one hand is
exactly solvable~\cite{MIGDAL} in the large $N$ expansion and on the other
hand may
have the property of inducing the gluon kinetic term. The reason of its
solvability must be found in the fact that the functional integration
over the angular degrees of freedom  can be exactly performed and after
that one is left with a "vector" model containing only the $N$ eigenvalues
of the scalar matrix that can be solved by a standard saddle point technique
determining the master field corresponding to the density of
eigenvalues.

Although it is not clear if this model or a modified version of it will
induce pure Yang-Mills theory it is, however, remarkable that one has
been able to construct matrix models where the sum over planar diagrams
can be explicitly performed in an arbitrary space-time dimension and not
just in zero and one dimension as in Ref~\cite{PARISI}.

In this paper we consider quantum chromodynamics in two dimensions
$(QCD_2)$  and we show that in the light cone gauge it is possible to
reformulate it completely in terms of a bilocal mesonic
field. The master field corresponding to the vacuum expectation value of
the bilocal mesonic field is then fixed in the limit of a large number of
colours by a saddle point equation whose solution is  equal to the fermion
propagator constructed in the original paper by 't Hooft~\cite{tH}.

We consider then the quadratic term containing the fluctuation around the
saddle point and we show that the equation of motion constructed from it
gives exactly the integral equation found in Ref.~\cite{tH} for the
mesonic spectrum.

Although our results are not new we feel that the method used could be
very useful for tackling more complicated problems that are still
unsolved.

\sect{The master field and the spectrum of QCD$_2$.}

We consider the action
\eq
S=\int d^2x~
\left\{
-{1\over 4g^2_0}tr(F^{\mu\nu}F_{\mu\nu})
+\pb^i(i\Dir-m^i_1\uno-m^i_2\gg_5)\pp^i
\right\}
\lbl{s-prima}
\en
where $F_{\mu\nu}=\part_\mu A_\nu-\part_\nu A_\mu +i[A_\mu,A_\nu]$,
$i\Dir_{A B}=\gg_\mu(i\part_\mu \uno_{A B}-A^a_\mu T^a_{A B})$,
$a,b=1...N^2-1$ are the indices of the adjoint representation of the
colour group, $A,B=1...N $ run over the fermionic representation
of the colour $SU(N)$ and $i,j...$ are flavour indices.
If we choose the gauge $A^a_-=0$ and we normalize the trace over the
fundamental representation to one, we can rewrite the previous
action\footnote{{\bf Conventions.}
$$
x^\pm=x_\mp=\usrd(x^0\pm x^1)~~~~
A^\mu B_\mu=
A_0 B_0 - A_1 B_1=
A_+ B_- + A_+ B_-
$$
$$
\gg_+=\gg^-=\mat{0}{\rd}{0}{0}~~~~
\gg_-=\gg^+=\mat{0}{0}{\rd}{0}~~~~
\gg_5=-\gg_0\gg_1=\mat{1}{0}{0}{-1}~~~~
P_{R,L}={1\pm\gg_5\over2}
$$
$$
\pp=\vett{\pp_+}{\pp_-}~~~~
\pb=\vet{\pb_-}{\pb_+}~~~~
$$
$$
\cc\pb=-\usrd
\mat
{\rd\pb P_R\cc}
{\pb \gg_-\cc}
{\pb \gg_+\cc}
{\rd\pb P_L\cc}
$$
}as
\eqa
S=\int&d^2x&
\biggl\{
{1\over 2g^2_0}(\dsi A^a_+)^2
+i\rd (\pbaip\dta\paip +\pbaim\dsi\paim)
\nonumber\\
&-&
(m_1^i+m_2^i)\pbaim\paip
-(m_1^i-m_2^i)\pbaip\paim
-A^a_+~\rd\pbaip T^a_{A B}\pbip
\biggr\}
\lbl{s-gauge}
\ena
Integrating over $A^a_+$ we get
\eqa
S&=&
\int d^2x~
\biggl\{
i\rd (\pbaip\dta\paip +\pbaim\dsi\paim)
-(m_1^i+m_2^i)\pbaim\paip
-(m_1^i-m_2^i)\pbaip\paim
\biggr\}
\nonumber\\
&&-g^2_0
\int d^2x~d^2y~
G(x-y)~
\pbaip(x) T^a_{A B}\pbip(x)~
\psibind{C}{j}{+}(y) T^a_{C D}\psiind{D}{j}{+}(y)=
\nonumber\\
&=&
\int d^2x~
\biggl\{
i\rd (\pbaip\dta\paip +\pbaim\dsi\paim)
-(m_1^i+m_2^i)\pbaim\paip
-(m_1^i-m_2^i)\pbaip\paim
\biggr\}
\nonumber\\
&+& g^2_0 x_R
\int d^2x~d^2y~
G(x-y)~
\biggl\{
\pbaip(x)\pajp(y)~
\psibind{B}{j}{+}(y) \pbip(x)
\nonumber\\&&~~~~
-{R\over N}\pbaip(x)\paip(x)~
\psibind{B}{j}{+}(y) \pbjp(y)
\biggr\}
\lbl{s}
\ena
where we used\footnote{
This relation is only valid for the fundamental representation of $SU(N)$, with
$x_R=1$ and $R=-1$.
We keep, however, $x_R$ and $R$ arbitrary in order to be able to trace back
easily the origin of the various terms that we get in the final expression.

Such a relation is essential for performing explicitly the large $N$ expansion,
as it will appear clearly in the following.
Its key property is that it allows one to rewrite the product of the two colour
currents as a product where the colour indices are saturated within each
current (such currents are colour singlets under the residual gauge
transformations for $x^+=y^+$).

}
$\sum_a T^a_{A B} T^a_{C D}=x_R(\dd_{B C} \dd_{A D}
+{R\over N}\dd_{A B} \dd_{C D})$ and
where $G(x)=-{1\over 2}\dd(x^+)\modu{x^-}=\ft{2}{k}{x}{1\over k_-^2}$.

The interaction term suggests to introduce the field
\eq
\rijmxy
=\sum_A \pbaj(y)~{\gg_-}~\pai(x)
\lbl{current1}
\en
and its partners
\eqa
\rijpxy
=\sum_A \pbaj(y)~{\gg_+}~\pai(x)
\nonumber\\
\sigma^{i j}(x,y)
=\sum_A \pbaj(y){\uno}~\pai(x)
\nonumber\\
\sigma_5^{i j}(x,y)
=\sum_A \pbaj(y)~{\gg_5}~\pai(x)
\nonumber\\
\sigma_{R,L}^{i j}(x,y)
= { {\sigma^{i j}(x,y) \pm \sigma_5^{i j}(x,y)}\over \rd}
\lbl{current2}
\ena

Now we want compute the jacobian of the transformation (see for instance
ref. \cite{DAS}) from the $\pb,\pp$ to the $\rr,\ss$.
\eq
\kern-3mm
\begin{array}{l}
J[\rhp,\rhm,\ss_R,\ss_L]=
\int[\bigd\pbaip ~\bigd\paip~\bigd\pbaim~\bigd\paim]
\\~~~~
\prod_{^{x y}_{i j}}
\dd[\rijmxy-\rd\sum_A \pbajp(y)\paip(x)]~
\prod_{^{x y}_{i j}}
\dd[\rijpxy-\rd\sum_A \pbajm(y)\paim(x)]
\\~~~~
\prod_{^{x y}_{i j}}
\dd[\sijpxy-\rd\sum_A \pbajm(y)\paip(x)]~
\prod_{^{x y}_{i j}}
\dd[\sijmxy-\rd\sum_A \pbajp(y)\paim(x)]
=
\\~
\\
=\int[\bigd\pbaip ~\bigd\paip~\bigd\pbaim~\bigd\paim]
[\bigd\aijp~\bigd\aijm~\bigd\bijp~\bigd\bijm]
\\~~~~
e^{
\ajimyx[\rijmxy-\rd\sum_A \pbajp(y)\paip(x)]~
+~\ajipyx[\rijpxy-\rd\sum_A \pbajm(y)\paim(x)]
}
\\~~~~
e^{
\bjipyx[\sijpxy-\rd\sum_A \pbajm(y)\paip(x)]~
+~\bjimyx[\sijmxy-\rd\sum_A \pbajp(y)\paim(x)]
}
\end{array}
\lbl{jacobian}
\en
where the sum over the flavour and space time indices is understood.

If we introduce the the matrices
\eq
\begin{array}{c}
M=||M_{P Q} ||=\mat{\bijpxy}{\aijpxy}{\aijmxy}{\bijmxy}
\\
U=||U_{P Q} ||=\mat{\sijpxy}{\rijmxy}{\rijpxy}{\sijmxy}
\\
  \begin{array}{cc}
  \PB^A_Q=\vet{\pbajm(y)}{\pbajp(y)} &
  \PP^A_P=\vett{\paip(x)}{\paim(x)}
  \end{array}
\end{array}
\en
where $P\equiv(x i \aa)$ and $Q\equiv(y j \bb)$,
we can rewrite the exponent of the integrand (\ref{jacobian}) as
\eqa
J[U]&=&\int [d \PB^A d\PP^A][dM]
\exp{[
Tr(MU)
-\rd\PB^A M\PP^A
]}
\nonumber\\
&\propto&\int [d M]
\exp{[
Tr(MU)
+N Tr \log M
]}
\ena
where $N$ is the dimension of the fermionic representation and
$Tr\equiv tr_x~tr_i~tr_\aa$.
Evaluating this integral with the saddle point method we get
\eq
J[U]\propto\exp[-N Tr\log U]
\en
where we have neglected non leading contribution in $N$.

If we define the matrix
\eqa
D=||D_{P Q}||&=&
\mat
{-\usrd(m_1^i+m_2^i)~\dd^{i j}~\dxy}
{i~\dd^{i j}~\part_{x^-}\dxy}
{i~\dd^{i j}~\part_{x^+}\dxy}
{-\usrd(m_1^i-m_2^i)\dd^{i j}\dxy}
=
\nonumber\\
&=&
\mat
{-m^{i j}_L\dxy}
{i~\dd^{i j}~\part_{x^-}\dxy}
{i~\dd^{i j}~\part_{x^+}\dxy}
{-m^{i j}_R\dxy}
\lbl{def-d}
\ena
and we rescale the master field $U\rightarrow N U$,  we can rewrite
the effective action as
\eqa
{1\over N}S_{eff}&=&
Tr(D U+i \log U)
+{1\over 2}g^2
\int d^2x~d^2y~G(x-y)~
U_{(x i 1),(y j 2)}~U_{(y j 1),(x i 2)}
\nonumber\\
&&-{1\over 2N}g^2R
\int d^2x~d^2y~G(x-y)~
U_{(x i 1),(x i 2)}~U_{(y j 1),(y j 2)}
\lbl{eff-s}
\ena
where $g^2=g^2_0 x_R N$.
Varying the effective action with respect  to $U_{Q P}$, we get the
equation for the master field, that reads to the leading order in $N$
\eq
D_{P Q}+i (U^{-1})_{P Q}
+g^2~\dd_{\aa,2}\dd_{\bb,1}
G(x-y)~
U_{(x i 2),(y j 1)}
=0
\en
Multiplying by $U$, we get immediately
\eq
D U_{P Q}+i \uno_{P Q}
+g^2~\dd_{\aa,2}
\int d^2z~
{}~G(x-z)
{}~U_{(x i 1),(z k 2)}
{}~U_{(z k 1), Q}
=0
\en
Writing explicitly these equations we find
\eq
\begin{array}{l}
 \left\{
 \begin{array}{l}
  i\part_{x^+}\rijmxy
  -m_R^{i l}\sljmxy
  +g^2\int d^2z~G(x-z)\rilmxz\rljmzy
  \\~~~~~
  +i~\dd^{i j}~\dxy
  =0
  \\
  i\part_{x^-}\sijmxy
  -m_L^{i l}\rljmxy
  =0
 \end{array}
 \right.
\\ \,
\\
 \left\{
 \begin{array}{l}
  i\part_{x^-}\rijpxy
  -m_L^{i l}\sljpxy
  +i~\dd^{i j}~\dxy
  =0
  \\
  i\part_{x^+}\sijpxy
  -m_R^{i l}\rljpxy
  +g^2\int d^2z~G(x-z)\rilmxz\sljpzy
  =0
 \end{array}
 \right.
 \end{array}
\lbl{propagatori}
\en
In particular if we eliminate $\sijmxy$ from the first equation using
the second one, we get the
fundamental equation
\eqa
i\part_{x^+}&\rijmxy&
  +i(m_R m_L)^{i l}~\int~d^2z~\dd(x^+-z^+)~\tt(x^--z^-)~\rljmzy
\nonumber\\
  &+&g^2\int d^2z~G(x-z)\rilmxz\rljmzy
  +i~\dd^{i j}~\dxy
=0
\lbl{prop}
\ena
In order to solve this equation it is better to pass to momentum space.
Since $\rijmxy=\rd <0|\sum_A \pbajp(y)\paip(x)|0>$ and the vacuum is
translationally invariant, we need only one momentum for the Fourier
transform of $\rijmxy$.
The previous equation (\ref{prop}) becomes
\eq
\left[
-p_+ \dd^{i k} +{(m_R m_L)^{i k}\over p_-}
+g^2\int d k~G(k)~\rr^{i k}_-(p-k)
\right]
\rr^{k j}_-(p)
+i \dij
=0
\lbl{prop2}
\en
and it suggests to set
\eqa
\rijmxy&=&
\ft{2}{p}{(x-y)}\rr^{i j}_-(p)=
\nonumber\\
&=&\dij \ft{2}{p}{(x-y)}
{2i~p_-\over 2p_+ p_- -2(m_R m_L)^i -p_- \GG(p)+i\ee}
\lbl{def-mom-r}
\ena

With this substitution eq. (\ref{prop2}) becomes eq. (\ref{def-d}) of
ref. \cite{tH}:
\eq
\GG(p)= {4g^2\over(2\pi)^2}\int {d^2k\over k_-^2}
{i(p_- + k_-)\over 2(p+k)_+(p+k)_- -2(m_R m_L)^i -(p+k)_- \GG(p+k)
+i\ee}
\lbl{def-GG}
\en
The explicit solution yields
\eq
\GG(p)=\GG(p_-) = {g^2\over \pi}
\left( {sgn(p_-)\over \lambda} - {1\over p_-} \right)
\en
where $\lambda$ is an infra-red cutoff as discussed in ref. \cite{tH}.

Inserting eq. (\ref{def-mom-r}) in eqs. (\ref{propagatori}), we get the
Fourier transform of the master field
\eq
U^{i j}_0(p)=
{i~\dd^{i j}\over 2p_+ p_- -2(m_R m_L)^i -p_- \GG(p)+i\ee}
\mat{-2 m_R^{i}}{2p_-}{2p_+ -\GG(p)}{-2 m_L^{i}}
\en
with

Eq. (\ref{def-GG}) determines the master field of QCD$_2$ that must be
identified with the vacuum expectation value of the quark propagator.

Now we consider the mass spectrum of the theory, i.e. the fluctuations
around the master field.
To this purpose we write $U=U_0+{1\over\sqrt{N}}\dd U$, and we
consider the terms in the effective action  that are $O(1)$ in $N$.
They are given by the quadratic terms in the fluctuation $\dd U$:
\eqa
S_{eff}^{(2)}&=&
-{i\over2}Tr(U_0^{-1}\dd U U_0^{-1}\dd U)
+{1\over 2}g^2
\int d^2x~d^2y~G(x-y)
{}~\dd U_{(x i 1),(y j 2)}
{}~\dd U_{(y j 1),(x i 2)}
\nonumber\\
&&-{g^2 R\over 2}\int d^2x~d^2y~G(x-y)U_{0(xi1),(xi1)}U_{0(y j2),(y j2)}
\lbl{s2}
\ena
The term
$-{g^2 R\over 2}\int d^2x~d^2y~G(x-y)U_{0(xi1),(xi1)}U_{0(y j2),(y j2)}$
is a constant and will be neglected.
If we compute $<\dd U_{P Q} ~\dd U_{R S}>$ from eq. (\ref{s2}) we find
eq. (7) of ref. \cite{CCG}, that describes the full quark-antiquark
scattering amplitude.
The spectrum of the theory is determined by the equation of motion for
the fields $\dd U$ that is given by
\eq
i~\dd U_{\aa\bb}^{i j}(x,y)
=
 g^2
\int d^2u~d^2v~
U_{0~~\aa,2}^{i k}(x-u)~
G(u-v)~\dd U_{1 2}^{k l}(u,v)~
U_{0~~1, \bb}^{l j}(v-y)
\en
and that in Fourier space leads to\footnote{
We define
$$\dd U(x,y)=\int {d^2 r\over (2\pi)^2}{d^2 s\over (2\pi)^2}
e^{ir\left({x+y\over2}\right)+is\left(x-y\right)}
\dd{\tilde U}(r,s)
$$
In the following we suppress the tilde over the Fourier transformed
fields.
}
\eq
\dd U^{i j}_{\aa\bb}(r,s) =
-i~g^2{T^{(i j)}_{\aa\bb} (s_-+{r_-\over 2}, s_--{r_-\over 2})
\over \Delta^{i}(s+{r\over 2}) \Delta^{j}(s-{r\over 2}) }
\int {d^2k\over (2\pi)^2} {1\over k_-^2 }~\dd U^{i j}_{1 2}(r,s-k)
\en
(no sum over $ i $ and $ j$ ), where
\eq
\Delta^i(p) = 2p_+p_- -2(m_R m_L)^{i} - p_-\GG(p) + i\epsilon
\en
 and
\eq
T^{(i j)}_{\aa\bb}(p_-,q_-) = \mat{4p_-m_R^{j}}{-4p_-q_-}
                                  {-4m_L^{i}m_R^{j}}{4m_L^{i}q_-}
\en

Following 't Hooft \cite{tH}, we integrate on both sides on $s_+$ and
 defining the gauge invariant field\footnote{
Notice that this is equivalent to set $x^+=y^+$ in $\dd U(x,y)$,
thus obtaining a gauge invariant object. If $x^+\ne y^+$ then $U(x,y)$
is not gauge invariant under the residual gauge transformations.
}
\eq
\vf^{i j}_{\aa\bb}(r,s_-) = \int {d s_+\over 2\pi}~ \dd U^{i j}_{\aa\bb}(r,s)
\en
we get choosing $r_->0$
\eqa
\vf^{i j}_{\aa\bb}(r,s_-) =
&&g^2
{ T^{(i j)}_{\aa\bb} (s_-+{r_-\over 2}, s_--{r_-\over 2})
  \over 4|s_-+{r_-\over 2}||s_--{r_-\over 2}| }
\left[ { M_i^2\over 2|s_-+{r_-\over 2}|}
     + { M_j^2\over 2|s_--{r_-\over 2}|}
     +{g^2\over\pi\lambda} - r_+ \right]^{-1}
\nonumber\\
&&\theta(s_-+{r_-\over 2}) \theta({r_-\over 2}- s_-)
\int {dk_-\over 2\pi k_-^2}~\vf^{i j}_{1 2}(r,s_--k_-)
\ena
where
\eq
M^2_i=2(m_R m_L)^i-{g^2\over\pi}
\en
In the sector $(\aa,\bb)=(2,1)$ it yields the 't Hooft starting equation
( eq. (15) of ref. \cite{tH}) when one identifies the Fourier transform of
$\rijmxy$ with $\pp(p,r)$.
In the other sectors requiring the cancellation of the IR cutoff $\ll$, we get
\eq
\vf^{i j}_{\aa\bb}(r,s_-) =
{T^{(i j)}_{\aa\bb} (s_-+{r_-\over 2}, s_--{r_-\over 2})
\over 4|s_-+{r_-\over 2}||s_--{r_-\over 2}| }
{}~\vf^{i j}_{1 2}(r,s_-)
\en
Performing the same straightforward manipulations as in ref. \cite{tH},
one is led to an
integral equation for the mass spectrum ( $\vf = \vf_{1 2} $;
we rescale $ s_- = r_-(x-{1\over2}) $ and define $\mu^2 = 2r_+r_-$ ):
\eq
\mu^2 \vf^{i j}(x) =
\left[{M_i^2\over x} +  {M_j^2\over (1-x)}\right]\vf^{i j}(x)
-{g^2\over\pi}~P\int_0^1 {\vf^{i j}(y)\over (y-x)^2}d y
\lbl{tH-eq}
\en
that is the famous 't Hooft equation, with a discrete spectrum of eigenvalues
labelled by an integer $n$ such that
$\mu^2_{n} \approx g^2\pi~n,~~ n\to \infty$.

In the other sectors we get the same equation for the mass spectrum, but
the mesonic fields change according to

\eq
\vf_{\aa\bb}^{i j}(x) = {\cal C}^{(i j)}_{\aa\bb}(x)\vf^{i j}(x)
\en
with
\eq
{\cal C}_{\aa\bb}^{(i j)} (x) =
\mat{{m^{j}_R \over (1-x) r_-}}{1}
    {-{m_L^{i}m_R^{j}\over x(1-x) r_-^2}}{-{m^{i}_L \over x r_-}}
\en

\sect{Conclusions}

We have shown that it is possible to derive the mass spectrum of QCD$_2$
in the large N limit with a functional approach. This approach allows
one to generalize the previous results in the presence of chiral masses.
The main point that guarantees the successful application of the large N
techniques is the possibility of defining colourless fields
(see eqs. (\ref{current1}) and (\ref{current2})) that allow one to extract
the $N$ dependence both in the action and in the measure of integration
(the mesonic bilocal $U(x,y)$ is a global colour singlet, but it is
variant under the residual gauge transformation unless $x^+=y^+$ !).

It would be interesting to discuss the $U(1)$ anomaly, the 3-meson
vertex and other "phenomenological" issues within this formalism.

{}~

One of us (P.DV) thanks D. Gross for a useful discussion.

\end{document}